\documentclass[a4paper,fleqn]{cas-sc}

\usepackage[authoryear,longnamesfirst]{natbib}

\usepackage{booktabs}
\usepackage{amsfonts}
\usepackage{amsmath}
\usepackage{amssymb}
\usepackage{nicefrac}
\usepackage{microtype}
\usepackage{xcolor}
\usepackage{graphicx}
\usepackage{multirow}
\usepackage{pifont}
\usepackage[section]{placeins}
\usepackage{url}

\newcommand{\threatforest}{\textsc{ThreatForest}}
\newcommand{\cmark}{\ding{51}}
\newcommand{\xmark}{\ding{55}}
\newcommand{\pmark}{$\sim$}

\begin{document}
\let\WriteBookmarks\relax
\def\floatpagepagefraction{1}
\def\textpagefraction{.001}

\shorttitle{\threatforest{}: Multi-Agent Attack Tree Generation}
\shortauthors{Leo et al.}

\title[mode=title]{\threatforest{}: Multi-Agent Attack Tree Generation with Pluggable TTP Framework Mapping}

\author[1]{Cristian Leo}[orcid=0009-0003-9269-8994]
\cormark[1]
\ead{crisleoo@amazon.com}
\credit{Conceptualization, Methodology, Software, Investigation, Formal analysis, Writing -- original draft, Writing -- review \& editing, Visualization}

\affiliation[1]{organization={Amazon Web Services, Inc.},
                addressline={410 Terry Avenue North},
                city={Seattle},
                state={WA},
                postcode={98109},
                country={United States}}

\author[2]{Anton Dykyi}
\ead{antondk@amazon.co.uk}
\credit{Methodology, Software, Validation, Writing -- review \& editing}

\affiliation[2]{organization={Amazon Web Services EMEA SARL, UK Branch},
                addressline={1 Principal Place, Worship Street},
                city={London},
                postcode={EC2A 2FA},
                country={United Kingdom}}

\author[2]{Danny Cortegaca}
\ead{dicorteg@amazon.co.uk}
\credit{Methodology, Investigation, Writing -- review \& editing}

\author[1]{Daniel Begimher}
\ead{dbbegimh@amazon.com}
\credit{Software, Validation, Writing -- review \& editing}

\author[1]{Prakash Jha}
\ead{prkasjh@amazon.com}
\credit{Conceptualization, Supervision, Project administration, Writing -- review \& editing}

\cortext[1]{Corresponding author. Address: Amazon Web Services, Inc., 410 Terry Avenue North, Seattle, WA 98109, United States. Email: \texttt{crisleoo@amazon.com}.}

\begin{abstract}
Threat modeling is essential for secure software development, yet manual analysis of cloud-native architectures is slow and demands scarce security expertise. We present \threatforest{}, a multi-agent system that generates structured attack trees from source code repositories, maps attack steps to adversary tactics, techniques, and procedures (TTPs) from a pluggable set of frameworks (MITRE ATT\&CK, CAPEC, and cloud-specific threat matrices), and synthesizes actionable mitigations. \threatforest{} decomposes threat modeling into a multi-stage agent pipeline---repository analysis, context refinement, threat generation, parallel attack-tree construction with TTP mapping and mitigation synthesis, and report generation---orchestrated as a directed graph with deterministic verification gates, bounded retries, and three human-in-the-loop validation points. A domain-specific sentence-transformer maps each attack step to candidate techniques by cosine similarity; we show empirically that this embedding stage, not the surrounding pipeline, is the dominant accuracy bottleneck. We evaluate \threatforest{} across seven application domains on a sixteen-dimension rubric, scored by a panel of independent LLM raters with an adversarial verification pass and human-in-the-loop expert review. Panel-measured quality reaches 0.63--0.68 (on a 0--1 scale) for threat statements, attack trees, and mitigations, but only 0.29 for embedding-only TTP mapping---a gap stable across all seven domains that isolates the binding constraint. A controlled single-call baseline on the same model more than doubles mapping defensibility, pinning the limitation on the embedding encoder rather than the multi-agent design. To our knowledge, \threatforest{} is the first end-to-end system that turns a code repository into TTP-mapped attack trees with evidence-based mitigations across multiple adversary frameworks, with a reusable evaluation framework for benchmarking such systems.
\end{abstract}


\maketitle

%==============================================================================
\section{Introduction}
\label{sec:intro}
%==============================================================================

Threat modeling is a cornerstone of secure software development~\citep{shostack2014threat}, requiring practitioners to systematically identify threats, analyze attack vectors, and propose mitigations. As cloud-native architectures grow in complexity---a single system may span dozens of managed services, each with distinct access control models, encryption configurations, and network boundaries---manual threat modeling becomes increasingly impractical. Industry surveys consistently report that organizations either skip threat modeling entirely or perform it only for the most critical systems, leaving significant attack surface unanalyzed~\citep{sotm2024}.

Existing tools address fragments of this problem. Diagram-driven tools such as OWASP Threat Dragon~\citep{owasp-threatdragon} provide structured diagramming but automate little of the core reasoning. AWS Threat Composer~\citep{aws-threatcomposer} offers guided threat statement authoring but requires manual input for each threat. Recent work has explored using large language models (LLMs) for threat identification~\citep{threatmodeling-llm2024,canllmsthreatmodel2025}, but these approaches typically stop at threat enumeration---they do not produce structured attack trees, map to standardized frameworks like MITRE ATT\&CK~\citep{strom2018mitre}, or generate actionable mitigations.

Attack trees, introduced by \citet{schneier1999attack}, provide a formal structure for modeling how an adversary achieves a goal through a hierarchy of sub-goals and concrete techniques. When combined with MITRE ATT\&CK technique mappings, attack trees become directly actionable: each leaf node maps to a known adversary behavior with documented detection and mitigation strategies. However, constructing these enriched attack trees manually requires deep expertise in both the target system and the threat landscape, and existing tools force a tradeoff between rigor (manual expert work) and coverage (running on every system, every release).

%----------------------------------------------------------------------
\subsection{Problem statement and scope}
\label{sec:problem}
%----------------------------------------------------------------------

We focus on the gap between \emph{threat identification} and \emph{actionable defensive guidance} for cloud-native applications. Concretely, given a source-code repository for a deployable application, we aim to automatically produce: (i) a set of high-level threats grounded in the application's specific technology stack and data flows; (ii) for each threat, a structured attack tree that decomposes the adversary's goal into AND/OR sub-goals and concrete leaf techniques; (iii) a mapping from each leaf technique to a known TTP from a configurable adversary framework (we treat MITRE ATT\&CK, CAPEC, and cloud-specific matrices as interchangeable backends); and (iv) for each TTP, a tailored mitigation that references the application's own components rather than generic advice.

\textbf{Threat model.} \threatforest{} is intended to operate in two cooperative postures with the user. In the \emph{advisory} posture the system runs autonomously over a repository the user owns, surfacing threats and suggested mitigations for downstream SME review. In the \emph{interactive} posture the system pauses at three human-in-the-loop gates (\S\ref{sec:hitl}) where SMEs validate or correct intermediate artifacts before they propagate. We assume the user has read access to the target repository and trusts the LLM provider; we do not assume the user is a security expert. Adversaries who target the modeled system are external; the analysis explicitly enumerates their capabilities, but \threatforest{} itself is not a defensive runtime control.

\textbf{Out of scope.} Runtime detection and response, automated patching, and offensive operations (e.g., autonomous penetration testing) are explicitly out of scope. \threatforest{} produces \emph{design-time} artifacts that feed into existing security review and SDL processes; it does not replace them. We also do not claim formal soundness of the resulting attack trees: the system is an LLM pipeline whose outputs require SME review for production use, and our evaluation framework (\S\ref{sec:evaluation}) is designed around exactly that workflow.

%----------------------------------------------------------------------
\subsection{Contributions}
\label{sec:contributions}
%----------------------------------------------------------------------

We present \threatforest{}, a multi-agent system that automates the threat modeling pipeline from source code repository to TTP-mapped attack trees with evidence-based mitigations across pluggable adversary frameworks. Our contributions are:

\begin{enumerate}
    \item \textbf{End-to-end multi-agent pipeline (\S\ref{sec:architecture}).} A ten-stage agent pipeline with human-in-the-loop validation gates that takes a code repository as input and produces structured attack trees, TTP mappings, and mitigations, orchestrated as a directed graph with deterministic verification gates and bounded retry semantics. The pipeline explicitly separates LLM agents from deterministic verifiers, and uses file-based state passing so every intermediate artifact is inspectable and resumable.

    \item \textbf{Framework-agnostic TTP mapping and isolation of the accuracy bottleneck (\S\ref{sec:ttp}, \S\ref{sec:ablation-embedding}).} A retrieval approach using ATTACK-BERT, a domain-specific sentence-transformer~\citep{attackbert}, maps LLM-generated attack steps to TTPs via cosine similarity, with top-$K$ candidates retained for downstream LLM- or SME-driven refinement and the framework swappable without retraining. Our central empirical finding is that this embedding stage---not threat generation, tree construction, or mitigation synthesis---is the dominant accuracy bottleneck: across all seven domains the rater panel judges only 29\% of ATTACK-BERT's mappings a defensible match, and general-purpose encoders recover even known-correct techniques far less often than ATTACK-BERT does. Isolating the bottleneck to a single, replaceable component tells future work precisely where to invest; as a pilot we show a cross-encoder reranker fine-tuned on our panel labels lifts mapping-judgment $F_1$ from 0.36 to 0.51.

    \item \textbf{Evaluation framework with a verified scoring protocol (\S\ref{sec:evaluation}).} A 16-dimension scoring rubric across four capabilities (threat statements, attack trees, TTP mapping, mitigations), wired through a Langfuse-based annotation queue~\citep{langfuse2024} that captures every agent interaction. We instantiate the rubric with a concrete scoring protocol---a panel of independent LLM raters plus an adversarial verifier, with human-in-the-loop SME adjudication---that scores every artifact across all seven domains (substantial inter-rater agreement: ordinal pairwise $0.93$, TTP Cohen's $\kappa = 0.78$), yielding measured rather than assumed quality numbers and a reproducible substrate for future systems to benchmark against.

    \item \textbf{Cross-domain empirical study with a single-call baseline (\S\ref{sec:results}, \S\ref{sec:ablation-monolithic}).} An evaluation across seven sample applications spanning IoT manufacturing, identity federation, generative AI, healthcare analytics, IAM governance, meeting transcription, and travel booking. Across all domains the system produces an average of 12 threats with 240 attack steps per application, maps to 89 unique techniques with full ATT\&CK tactic coverage in five of seven domains, and attains panel-measured quality of 0.63--0.68 for threat statements, attack trees, and mitigations versus 0.29 for embedding-only TTP mapping---a gap stable across domains. A controlled monolithic single-call baseline on the same model isolates what the decomposition buys: the pipeline wins decisively on coverage and structural uniformity, while a single call matches or exceeds it on per-item quality and, tellingly, more than doubles TTP-mapping defensibility (0.63 vs.\ 0.29)---confirming the bottleneck is the embedding encoder, not the architecture.
\end{enumerate}

The remainder of the paper is organized as follows. Section~\ref{sec:related} situates the work against existing threat-modeling tools, LLM-based security analysis, and ATT\&CK mapping literature. Section~\ref{sec:architecture} describes the pipeline and HITL gates in detail. Section~\ref{sec:evaluation} defines the evaluation framework. Section~\ref{sec:results} reports cross-domain results. Section~\ref{sec:discussion} discusses design tradeoffs, threats to validity, and future work.

%==============================================================================
\section{Related Work}
\label{sec:related}
%==============================================================================

We organize prior work into four areas: classical threat-modeling tools and methodologies, LLM-based threat analysis and offensive reasoning, automated MITRE ATT\&CK mapping, and the supporting infrastructure for domain-specific encoders and multi-agent systems. \threatforest{} sits at the intersection of all four: it inherits the structured-output discipline of classical threat modeling, leverages LLM agents for the generation steps that previously required SME expertise, uses embedding-based retrieval to bridge generated text to a standardized technique catalog, and is implemented on top of a graph-orchestration framework.

%----------------------------------------------------------------------
\subsection{Threat modeling tools and methodologies}
%----------------------------------------------------------------------

STRIDE~\citep{shostack2014threat} and attack trees~\citep{schneier1999attack} provide the methodological foundation for systematic threat analysis: STRIDE enumerates threat categories (Spoofing, Tampering, Repudiation, Information disclosure, Denial of service, Elevation of privilege) over a data-flow diagram, while attack trees decompose adversary goals into AND/OR sub-goals. Both predate widespread cloud-native deployment and assume a security expert manually constructs the model for each system. The cost of that assumption has driven a generation of tool-assisted approaches.

Diagram-driven threat-modeling tools---of which OWASP Threat Dragon~\citep{owasp-threatdragon} is a representative open-source example---offer drag-and-drop data-flow diagramming with rule-based threat suggestion: the tool surfaces canned STRIDE threats based on element types and trust-boundary crossings. AWS Threat Composer~\citep{aws-threatcomposer} provides a guided authoring workflow for individual threat statements. None of these tools take source code as input, none produce attack trees beyond a flat threat list, and none map outputs to a standardized technique framework. They are useful but require the practitioner to do the underlying threat reasoning manually.

\threatforest{} is positioned as a complement rather than a replacement: practitioners can feed its output into the same review processes those tools support, but the generation step is automated and the output is structured for downstream consumption.

%----------------------------------------------------------------------
\subsection{LLM-based threat analysis and offensive reasoning}
%----------------------------------------------------------------------

A recent line of work applies LLMs directly to security tasks. ThreatModeling-LLM~\citep{threatmodeling-llm2024} fine-tunes models for threat identification in banking systems, and \citet{canllmsthreatmodel2025} systematically evaluate whether frontier LLMs can identify threats in real-world cloud infrastructure, finding that performance varies substantially across prompting strategies. These works focus on the threat-\emph{identification} step alone---they stop where \threatforest{} continues, and do not produce attack trees, TTP mappings, or mitigations.

A parallel strand evaluates LLMs in offensive roles. PentestGPT~\citep{deng2024pentestgpt} and Happe and Cito~\citep{happe2023getting} treat the LLM as an autonomous penetration tester that interacts with a target system, while Cybench~\citep{zhang2024cybench} provides a CTF-style benchmark for cyber-reasoning capabilities and SecureFalcon~\citep{ferrag2023securefalcon} explores LLMs for vulnerability reasoning over source code. Although these systems share the LLM-agent substrate, their objectives are different: they exercise the model's offensive capability on a live target, whereas \threatforest{} produces a design-time artifact for defensive use. The structural-validity requirements (every output JSON must be schema-conformant; every leaf node must reference a real component) and the HITL gate design are tailored to the defensive setting.

%----------------------------------------------------------------------
\subsection{Automated MITRE ATT\&CK mapping}
%----------------------------------------------------------------------

Mapping unstructured security text to MITRE ATT\&CK has been studied extensively, almost always as a classification problem over CTI reports or CVE descriptions. CVE2ATT\&CK~\citep{cve2attack2022} fine-tunes BERT for technique classification on CVE summaries, and SMET~\citep{smet2023} uses Siamese-network embeddings for the same task. \citet{modernbert2025} combine ModernBERT with BERTopic to surface tactic-level clusters, and earlier work---TTPDrill~\citep{husari2017ttpdrill} and rcATT~\citep{legoy2020rcatt}---extracts technique mentions from CTI reports using rule-based and shallow NLP techniques. \citet{xiong2022cyberkg} build a graph-based threat model with ATT\&CK as a knowledge backbone for risk analysis.

These works share an important assumption that does \emph{not} hold for \threatforest{}: their input is an existing security text (a CVE summary, an incident report) where the underlying technique is grounded by prior reporting. \threatforest{} maps \emph{generated} attack-tree steps written by an LLM, so the ground truth is itself uncertain and the retrieval problem is harder---we are not classifying a known-good description, we are matching a model's hypothesized attack step against a structured technique catalog. Our results (\S\ref{sec:results}) confirm that off-the-shelf encoders struggle with this regime, and motivate the specialized threat-encoder fine-tuning we describe as future work.

%----------------------------------------------------------------------
\subsection{Domain-specific encoders, multi-agent systems, and orchestration}
%----------------------------------------------------------------------

Domain-specific text encoders for cybersecurity have proliferated in recent years. SecureBERT~\citep{aghaei2023securebert} adapts BERT to cybersecurity text, and ATTACK-BERT~\citep{attackbert} extends sentence-transformers~\citep{reimers2019sbert} with attack-action embeddings; we use the latter as our retrieval backbone (\S\ref{sec:ttp}). Beyond classification and retrieval, recent work has explored multi-agent orchestration as a substrate for complex LLM applications: open-source frameworks such as Strands Agents~\citep{strands2025} and LangGraph~\citep{langgraph2024} provide graph-based coordination of multiple agents with explicit retry and verification semantics, building on prompting and tool-use techniques like Chain-of-Thought~\citep{wei2022cot}, ReAct~\citep{yao2023react}, and Toolformer~\citep{schick2023toolformer}.

Multi-agent systems have been applied to code generation, automated reasoning, and recently to data analysis and software engineering, but their use for \emph{security threat modeling}---where outputs must be both technically accurate and structurally valid in ways that downstream tooling can consume---remains largely unexplored. \threatforest{} is, to our knowledge, the first end-to-end multi-agent system for this task that combines pluggable TTP framework mapping, deterministic verification gates, and human-in-the-loop SME review in a single pipeline.

Finally, our evaluation builds on the LLM-as-a-judge paradigm, in which a strong model scores another model's output in place of costly human annotation~\citep{zheng2023judging,gu2024surveyjudge}. That literature also documents the failure modes we design against---judges are prone to leniency and self-preference bias and exhibit only moderate agreement with human raters~\citep{gu2024surveyjudge}. We mitigate these with a multi-rater panel, an adversarial verifier that may only lower scores, and a cross-model calibration check (\S\ref{sec:results}); the calibration in fact shows our adversarial panel is \emph{more} conservative than a single-pass judge, directly countering the leniency bias that paradigm is known for.

%==============================================================================
\section{System Architecture}
\label{sec:architecture}
%==============================================================================

\begin{figure}[!htbp]
\centering
\includegraphics[width=\textwidth]{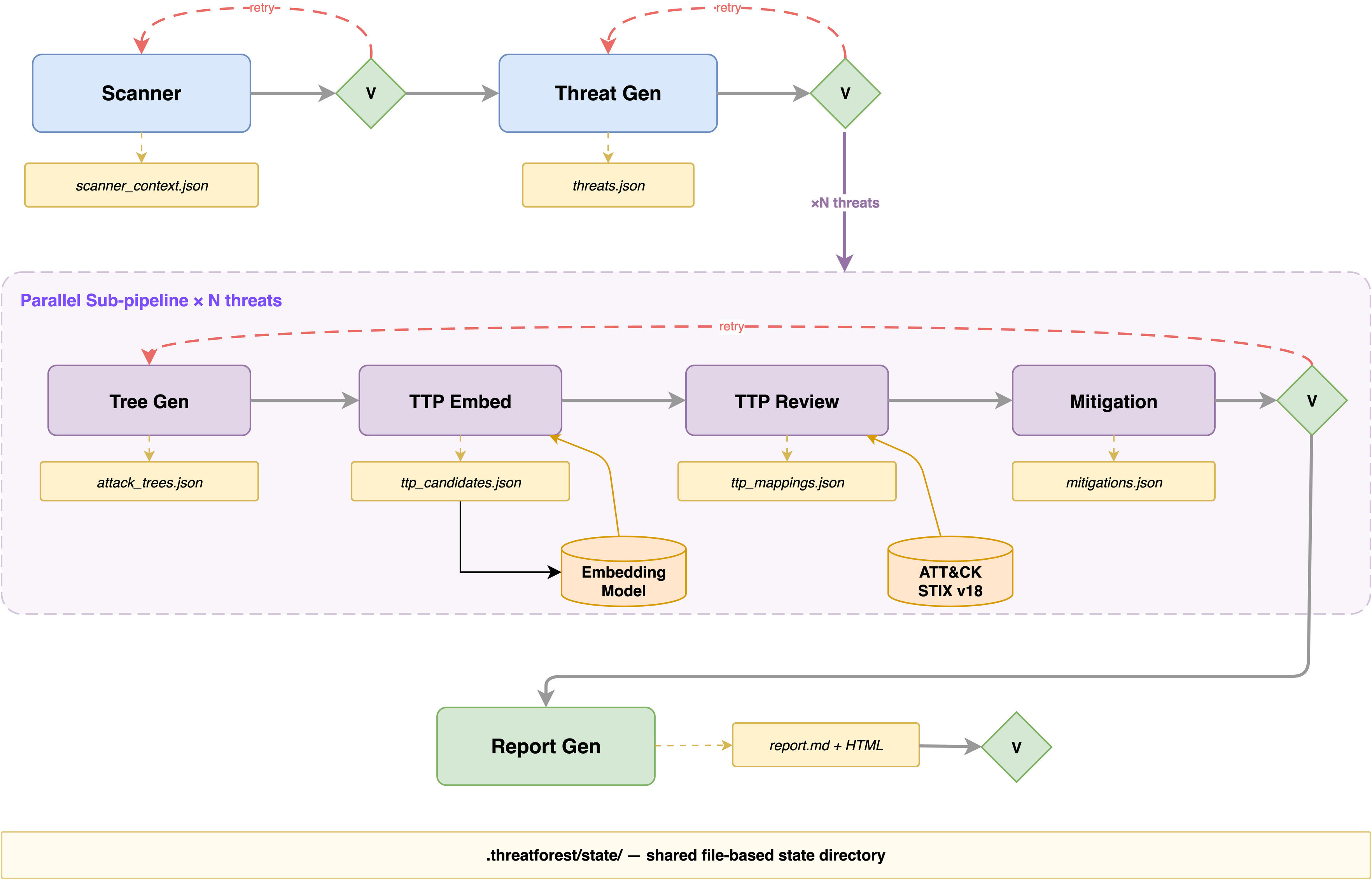}
\caption{The \threatforest{} pipeline. Rounded boxes are LLM agents; diamonds are deterministic verifiers; shaded boxes are human-in-the-loop (HITL) review gates; yellow boxes are JSON state files written to the shared \texttt{.threatforest/state/} directory. Dashed red arrows indicate retry edges on verification failure. After the scanner, a scanner-review gate and an interviewer agent refine the project context before threat generation; a threat-review gate triages threats before the parallel fan-out. The tree$\to$TTP$\to$mitigation sub-pipeline (purple region) runs in parallel for each threat. A deterministic probability stage then annotates each step before the report generator. The embedding model and ATT\&CK STIX knowledge graph feed into the TTP stages.}
\label{fig:pipeline}

\noindent\textit{Alt text:} A directed-graph diagram of the \threatforest{} pipeline. Left-to-right boxes show the stages: Scanner, Scanner Review, Interviewer, Threat Generation, Threat Review, then a Parallel Sub-pipeline (Tree Generation, TTP Embedding, TTP Review, Mitigation) repeated per threat, then a Probability stage and a Report Generator. Diamond verifier nodes follow each LLM agent. Yellow cylinders represent JSON state files (scanner\_context.json, threats.json, attack\_trees.json, ttp\_mappings.json, mitigations.json, report.md). Dashed red arrows from each verifier back to its producing agent indicate retry edges. Side panels show the embedding model and ATT\&CK STIX knowledge graph feeding into the TTP-mapping stage.
\end{figure}

\threatforest{} decomposes threat modeling into ten agent stages plus verification gates, each implemented as an autonomous agent with tool access, orchestrated as a directed graph. Figure~\ref{fig:pipeline} shows the full pipeline.

\paragraph{Scope of the contribution.}
The individual techniques on which \threatforest{} relies---LLM agents with tool use, sentence-transformer retrieval, graph-based multi-agent orchestration, deterministic schema-validated verifiers, human-in-the-loop interrupt gates---are all established in prior work. We do not claim novelty for any single component. The contribution is the \emph{composition}: to our knowledge, no prior system combines these components into an end-to-end pipeline that takes a source-code repository as input and produces TTP-mapped attack trees, evidence-grounded mitigations, and probability-annotated reach paths as a single artifact (Section~\ref{sec:comparison} compares capability coverage against the closest representative tools). We claim three things empirically: (i) this composition produces structurally valid output across seven heterogeneous cloud-native architectures (Section~\ref{sec:results}); (ii) the embedding-based TTP retrieval stage is the dominant accuracy bottleneck and is bounded above by the encoder choice rather than any other pipeline parameter (Section~\ref{sec:ablation-embedding}); and (iii) the resulting evaluation framework is reusable as a baseline against which future systems can be measured.

%----------------------------------------------------------------------
\subsection{Pipeline Overview and Graph Orchestration}
\label{sec:orchestration}
%----------------------------------------------------------------------

The pipeline is implemented as a Strands Graph~\citep{strands2025}: nodes are agent executors and edges carry conditional predicates. Each agent reads its predecessor's output from a shared state directory (\texttt{.threatforest/state/}) and writes structured JSON, providing inspectable intermediate outputs, resumability after failures, and parallel-safe execution. The graph defines retry edges---when a verifier rejects an output, execution flows back to the producing agent for a second attempt. To prevent unbounded retries we impose a budget $B \geq |V| + k \cdot |\{\text{retriable stages}\}|$ on total node executions, where $|V|$ is the happy-path node count and $k$ is the maximum retries per stage. In our configuration, $|V| = 12$, $k = 2$, and four stages carry retry edges (scanner, threat, parallel, report), yielding $B \geq 20$; we set the runtime ceiling to $B = 32$ to absorb rare multi-stage retry chains. Any execution that exhausts the budget terminates with a partial result rather than looping indefinitely.

%----------------------------------------------------------------------
\subsection{Scanner and Threat Generation}
\label{sec:scanner-threat}
%----------------------------------------------------------------------

\textbf{Scanner Agent.} The scanner receives a repository path and uses two tools: a \emph{structural analyzer} that extracts the project's file tree, dependency manifests, and infrastructure-as-code definitions; and a \emph{sandboxed file reader} restricted to the repository directory. The agent produces a \emph{project context} containing: cloud provider, technology stack, services, authentication mechanisms, data flows, and deployment model. A deterministic verifier checks that all required fields are present and non-empty.

\textbf{Threat Agent.} The threat agent receives the project context and generates structured threat statements following the pattern:

\begin{quote}
\emph{A [actor] with [access/knowledge], can [action], which leads to [impact], resulting in [consequence].}
\end{quote}

Each threat includes a title, description, priority, and affected components. The verifier validates JSON structure and ensures each threat references components identified in the scanner context.

%----------------------------------------------------------------------
\subsection{Human-in-the-Loop Validation Gates}
\label{sec:hitl}
%----------------------------------------------------------------------

Three interactive gates let subject-matter experts (SMEs) correct pipeline output before it propagates downstream. Each gate reads the upstream state file, emits a structured interrupt to the UI, and applies the user's response: deterministic edits are merged directly into state files, while free-text feedback re-invokes the upstream LLM agent.

\textbf{Scanner Review} surfaces the inferred project context (cloud provider, technology stack, services, authentication mechanisms, data sensitivity, compliance) for confirmation or JSON edits. \textbf{Interviewer Agent} pairs a fixed opening question (deployment stage) with a tool-equipped LLM agent that asks targeted follow-ups via an \texttt{ask\_user} tool and enriches the context with a confidence score and summary; users can loop back to scanner review via a \texttt{\_\_back\_\_} sentinel. \textbf{Threat Review} presents each threat with four guided questions (sensibility, prioritization, false positives, missing threats); structured edits (priority changes, removals, reordering) apply deterministically to \texttt{threats.json}, while free-text feedback re-invokes the threat agent. The gate loops until the user sends \texttt{proceed}.

These gates are the primary mechanism by which SME knowledge enters the pipeline and supply high-quality signal for the evaluation infrastructure (\S\ref{sec:evaluation}).

%----------------------------------------------------------------------
\subsection{Attack Tree Generation}
\label{sec:tree}
%----------------------------------------------------------------------

Attack trees are generated in parallel: one sub-pipeline per threat, all executing concurrently. Each tree agent receives the scanner context and a single threat and produces a hierarchical tree whose root represents the adversary's goal, intermediate nodes decompose it into AND/OR sub-goals, and leaves are concrete attack techniques. Formally, an attack tree for threat $\theta$ is a rooted directed tree $\mathcal{T}_\theta = (S, E, r)$ with steps $S = \{s_1, \ldots, s_n\}$, parent-child edges $E \subseteq S \times S$, and root $r$. Each step $s_i = (\textit{id}, \textit{title}, \textit{description}, \textit{parent\_id}, \textit{is\_leaf}, \textit{category})$ where $category \in \{\text{fact}, \text{action}, \text{detection}, \text{gate}\}$ classifies its role (preconditions, adversary operations, observable indicators, AND/OR gates). We measure tree complexity by depth---the longest root-to-leaf path:
\begin{equation}
\text{depth}(\mathcal{T}_\theta) = \max_{\ell \in \text{leaves}(S)} \, |\text{path}(r, \ell)|
\label{eq:depth}
\end{equation}
The set of attack paths $\mathcal{P}_\theta = \{\text{path}(r, \ell) \mid \ell \in \text{leaves}(S)\}$ enumerates complete attack scenarios; the union $\bigcup_\theta \mathcal{P}_\theta$ across threats defines the modeled attack surface. For $N$ threats, the parallel fan-out produces $N$ independent sub-pipelines (tree $\to$ TTP $\to$ mitigation), providing up to $N\times$ speedup.

%----------------------------------------------------------------------
\subsection{TTP Mapping via Embedding Retrieval}
\label{sec:ttp}
%----------------------------------------------------------------------

We map attack-tree steps to MITRE ATT\&CK techniques~\citep{mitre-attack} via embedding-based retrieval. We benchmarked several general-purpose and domain-specific sentence-transformers---E5-large-v2~\citep{wang2022e5}, all-mpnet-base-v2~\citep{mpnet2021}, BGE-large-en-v1.5~\citep{xiao2024bge}, and Qwen3-Embedding-0.6B~\citep{qwen3embedding2025}---and selected ATTACK-BERT~\citep{attackbert} for its competitive accuracy and smaller footprint (768-d embeddings, fast CPU inference). All five models are released under permissive open-source licenses (Apache-2.0 or MIT) and are used here under the terms of those licenses. Given an attack tree with $n$ steps and the ATT\&CK matrix of $M$ technique descriptions, we encode each step description $d_i$ and measure cosine similarity to every technique $t_j$:
\begin{equation}
\text{sim}(d_i, t_j) = \frac{\mathbf{e}(d_i) \cdot \mathbf{e}(t_j)}{\|\mathbf{e}(d_i)\| \, \|\mathbf{e}(t_j)\|}
\label{eq:cosine}
\end{equation}
where $\mathbf{e}(\cdot): \text{text} \to \mathbb{R}^{768}$ is the ATTACK-BERT embedding. For each step we retrieve the top-$K$ candidates with $\text{sim} \geq \tau$ and select the top-1 as the final mapping; we set $K=3$ and $\tau=0.3$ empirically. Steps where no candidate exceeds $\tau$ receive no mapping, avoiding forced low-confidence assignments. The remaining $K$ candidates are retained alongside each mapping to support downstream refinement via LLM-based review or SME validation (\S\ref{sec:evaluation}).

%----------------------------------------------------------------------
\subsection{Mitigation and Verification}
\label{sec:mitigation}
%----------------------------------------------------------------------

\textbf{Mitigation Agent.} For each threat $\theta_i$, the mitigation agent is conditioned on three inputs---scanner context $\mathcal{C}$, attack tree $\mathcal{T}_{\theta_i}$, and TTP mappings $\mathcal{M}_{\theta_i} = \{(s_j, t_j)\}$---and produces one mitigation per unique mapped technique: $\text{mitigate}(\theta_i) = f_{\text{LLM}}(\mathcal{C}, \mathcal{T}_{\theta_i}, \mathcal{M}_{\theta_i}) \to \{m_1, \ldots, m_p\}$. Each $m_k$ is a structured record (\emph{technique\_id}, \emph{guidance}, \emph{priority}, \emph{evidence}, \emph{type}) validated at tool-call time by a Pydantic schema. The \emph{type} field classifies remediation urgency (quick win, short-term, medium-term, long-term, monitoring), and \emph{evidence} links each mitigation back to the attack steps and techniques that motivate it. Conditioning on $\mathcal{C}$ grounds mitigations in concrete infrastructure rather than generic advice; on $\mathcal{T}_{\theta_i}$ enables reasoning over multi-step chains; and on $\mathcal{M}_{\theta_i}$ enables coverage verification---the per-threat verifier confirms that every mapped technique has a corresponding mitigation.

\textbf{Verifier Pattern.} Every stage in the pipeline includes a deterministic verifier---a pure function (no LLM) that validates structural properties of the output:

\begin{itemize}
    \item \textbf{Scanner verifier}: All required context fields present and non-empty.
    \item \textbf{Threat verifier}: Valid JSON, each threat has required fields, references valid components.
    \item \textbf{Pipeline verifier}: Attack trees have valid parent-child relationships, TTP mappings reference existing steps, mitigations reference existing techniques.
    \item \textbf{Per-threat mitigation coverage}: Within the parallel sub-pipeline each threat is independently checked for mitigation coverage. Structural failures (missing fields, invalid references) hard-fail and trigger retry; coverage gaps are emitted as soft warnings, avoiding expensive full-pipeline retries for minor omissions.
    \item \textbf{Report verifier}: Output file exists with expected sections.
\end{itemize}

On verification failure, the graph's retry edge routes execution back to the producing agent. This pattern provides quality gates without the cost or non-determinism of LLM-based validation.

%----------------------------------------------------------------------
\subsection{Probability Scoring}
\label{sec:probability}
%----------------------------------------------------------------------

Between the parallel sub-pipeline and the report generator, a pure-Python probability stage annotates every attack step with a success probability. A per-step prior from step attributes (category, tree depth, TTP mapping presence) is updated via a Bayesian posterior driven by the TTP similarity score, mitigation priority, feasibility notes, and a tech-stack mismatch detector that lowers probability when a step references technology absent from the scanner context. Fact-category steps (preconditions) are pinned to $1.0$. A Markov rollup then propagates per-step probabilities along tree edges to produce a \emph{reach probability}---the likelihood an adversary reaches each node given AND/OR gate semantics. The stage is deterministic and idempotent, so it carries no retry edge.

\textbf{Report Generator.} The final stage is deterministic (no LLM): it compiles all state files into a structured Markdown report and an interactive HTML dashboard with attack tree visualizations. Probability-annotated trees surface the highest-reach paths first.

%==============================================================================
\section{Evaluation Methodology}
\label{sec:evaluation}
%==============================================================================

We design an evaluation framework that combines automated structural metrics with human expert assessment, supported by an observability infrastructure that captures every agent interaction. The seven application domains used in the empirical study (Section~\ref{sec:results}) are deliberately chosen to span common cloud-native architecture archetypes---IoT manufacturing, identity federation, generative AI, healthcare analytics, IAM governance, meeting transcription, and travel booking---so that the cross-domain numbers function as a test of external validity rather than a single-domain proof of concept.

%----------------------------------------------------------------------
\subsection{Scoring Dimensions}
%----------------------------------------------------------------------

We define 16 scoring dimensions across four capabilities (Table~\ref{tab:scoring-dims}), each using a 5-point categorical scale
\begin{equation*}
\begin{aligned}
\mathcal{C} = \{&\text{excellent}, \text{good}, \text{acceptable}, \\
                &\text{poor}, \text{unacceptable}\}
\end{aligned}
\end{equation*}
mapped to numeric values $\{1.0, 0.75, 0.5, 0.25, 0.0\}$. TTP Mapping uses a binary accuracy scale where an SME (or LLM judge) labels each mapping as \emph{good}~(1) or \emph{bad}~(0).

\begin{table}[t]
\centering
\small
\caption{Scoring dimensions for \threatforest{} evaluation. All use the 5-point categorical scale unless noted.}
\label{tab:scoring-dims}
\begin{tabular}{llp{9cm}}
\toprule
\textbf{Capability} & \textbf{Dimension} & \textbf{Description} \\
\midrule
\multirow{5}{*}{Threat Statements}
  & Overall Quality & Holistic assessment of generated threats \\
  & Relevance to Context & Match to application context \\
  & Completeness & Coverage of threat categories \\
  & Technical Accuracy & Technical correctness \\
  & Hallucination Detection & Absence of fabricated content \\
\midrule
\multirow{6}{*}{Attack Trees}
  & Overall Quality & Holistic assessment of the attack tree \\
  & Structural Quality & Depth, branching, organization \\
  & Technical Realism & Feasibility of attack techniques \\
  & Attack Path Logic & Logical progression from access to impact \\
  & Completeness & Coverage of attack vectors and phases \\
  & Actionability & Usefulness for defenders \\
\midrule
TTP Mapping
  & Mapping Accuracy & Quality of MITRE ATT\&CK technique mapping$^*$ \\
\midrule
\multirow{4}{*}{Mitigations}
  & Actionability & Concrete, implementable steps \\
  & Specificity & Tailored to identified threat and tech stack \\
  & Coverage & Addresses all identified attack paths \\
  & Technical Accuracy & Correctness of recommended controls \\
\bottomrule
\multicolumn{3}{l}{\footnotesize $^*$Binary accuracy: each mapping judged as good (1) or bad (0).}
\end{tabular}
\end{table}

%----------------------------------------------------------------------
\subsection{Tracing and Review Infrastructure}
%----------------------------------------------------------------------

\threatforest{} captures two complementary trace types via Langfuse~\citep{langfuse2024}: \emph{OTEL traces} emitted automatically by the Strands SDK~\citep{strands2025} for every LLM call, tool invocation, and event loop cycle, and \emph{annotation traces} pushed at each subgraph boundary with clean input/output JSON. Annotation traces are routed to per-capability Langfuse queues where reviewers assign categorical scores independently for each dimension; individual TTP mappings are also pushed as dataset items for building a ground truth corpus that supports precision@$K$ evaluation and fine-tuning.

%----------------------------------------------------------------------
\subsection{Scoring protocol: LLM rater panel with SME verification}
\label{sec:scoring-protocol}
%----------------------------------------------------------------------

Scoring every artifact in the study by hand is costly, so we adopt a two-stage protocol that combines an automated panel with human-in-the-loop verification. For each artifact (threat statement, attack tree, TTP mapping, mitigation), a panel of \emph{three independent large-language-model rater agents} assigns a categorical label on every applicable dimension of Table~\ref{tab:scoring-dims}, conditioned on the same scanner context the pipeline used, so that context-dependent dimensions (relevance, specificity) are judgeable. The three raters are given deliberately different review foci---one scores critically and independently, one is instructed to penalize content not grounded in the application's declared stack, and one attends to completeness and hallucination---to diversify failure detection rather than merely replicate one viewpoint. An \emph{adversarial verifier agent} then audits the three rater outputs against the artifact: it starts from the per-dimension median and may only \emph{confirm or lower} a score, lowering when it can cite a concrete defect (generic or boilerplate content, a technical error, a hallucinated component, or missing coverage; for TTP mapping it marks a label \emph{bad} whenever the assigned technique is not a defensible match for the attack step). The reconciled labels are the panel's output.

This panel is a \emph{first-pass} evaluator. Its reconciled labels and per-item rationales are then routed to the Langfuse annotation queues for \emph{human SME adjudication}: an SME reviews the panel's judgments and the cited defects and confirms or corrects them. We report panel-derived scores as the primary quantitative signal because they cover every artifact in the study; the human SME pass is a confirmatory review over the same annotation queues, not a precondition for the reported numbers (which are panel measurements, explicitly not unaided human annotation). To quantify the panel's reliability we report two checks in \S\ref{sec:results}: inter-rater agreement on the raw (pre-verifier) rater labels (mean pairwise agreement for the ordinal dimensions; Cohen's $\kappa$ and percent agreement for the binary TTP dimension), and a cross-model calibration in which an independent judge on a different LLM family re-scores a stratified sample blind to the panel.

%----------------------------------------------------------------------
\subsection{Automated Structural Metrics}
%----------------------------------------------------------------------

We compute the following metrics automatically from pipeline outputs. Let $\mathcal{A} = \{\mathcal{T}_1, \ldots, \mathcal{T}_N\}$ denote the set of attack trees for an application, $\mathcal{M} = \{m_1, \ldots, m_n\}$ the set of TTP mappings, and $\tau(m_j)$ the technique assigned to mapping $m_j$.

\textbf{Tree depth} is defined per Eq.~\ref{eq:depth}; we report the mean across trees: $\bar{D} = \frac{1}{N} \sum_{i=1}^{N} \text{depth}(\mathcal{T}_i)$.

\textbf{Technique diversity} measures the breadth of ATT\&CK coverage:
\begin{equation}
\delta = \frac{|\{\tau(m_j) \mid j = 1, \ldots, n\}|}{n}
\label{eq:diversity}
\end{equation}
where $\delta = 1$ indicates all mappings are unique and $\delta \to 0$ indicates heavy reuse of a few techniques.

\textbf{Phase coverage} measures the fraction of the 14 ATT\&CK tactics $\Phi = \{\phi_1, \ldots, \phi_{14}\}$ (Reconnaissance through Impact) represented. Let $\text{tactic}(t)$ return the set of tactics associated with technique $t$:
\begin{equation}
\gamma = \frac{\left|\bigcup_{j=1}^{n} \text{tactic}(\tau(m_j))\right|}{|\Phi|}
\label{eq:phase-coverage}
\end{equation}

\textbf{Aggregate quality score.} For each application $a$ and scoring dimension $d$, let $q_{a,d} \in \{0, 0.25, 0.5, 0.75, 1.0\}$ be the assigned score. We report the mean and standard deviation across applications:
\begin{equation}
\begin{aligned}
\bar{q}_d &= \frac{1}{|\mathcal{A}|} \sum_{a} q_{a,d}, \\
\sigma_d &= \sqrt{\frac{1}{|\mathcal{A}|} \sum_{a} (q_{a,d} - \bar{q}_d)^2}
\end{aligned}
\label{eq:agg-score}
\end{equation}

%==============================================================================
\section{Results}
\label{sec:results}
%==============================================================================

We evaluate \threatforest{} on 7 sample applications spanning diverse cloud architectures: IoT manufacturing, identity federation, generative AI, healthcare analytics, IAM governance, meeting transcription, and travel booking. All experiments use Claude Sonnet~4.5 (Anthropic) served via Amazon Bedrock at temperature~0,\footnote{Bedrock model ID \texttt{global.anthropic.claude-sonnet-4-5-20250929-v1:0}; all agents run at temperature~0 except the interviewer follow-up agent (0.3).} with ATTACK-BERT \citep{attackbert} for TTP embedding retrieval against the MITRE ATT\&CK STIX dataset \citep{mitre-attack}.

%----------------------------------------------------------------------
\subsection{A worked example}
\label{sec:worked-example}
%----------------------------------------------------------------------

Before the aggregate numbers, Table~\ref{tab:worked} traces one threat end-to-end through the pipeline for the generative-AI chatbot application, showing the joint artifact a single run produces and---on one leaf---the embedding-mapping failure the rest of this section quantifies.

\begin{table}[t]
\centering
\footnotesize
\caption{One threat from the GenAI Chatbot domain traced through the pipeline (abbreviated). The TTP column shows ATTACK-BERT's top-1 with the panel's verdict: the precondition leaf is mismapped to \emph{Serverless Execution} (a weak match the panel marks \xmark), while the scan-and-extract leaf maps defensibly to \emph{Chat Messages} (\cmark)---the per-leaf coin-flip that yields the 0.29 aggregate.}
\label{tab:worked}
\begin{tabular}{p{0.97\columnwidth}}
\toprule
\textbf{Threat (TS002, priority high).} ``A malicious insider with IAM permissions to DynamoDB, or a compromised Lambda execution role, can query the chat-history table containing full conversation logs, leading to unauthorized access to PII not redacted by Bedrock Guardrails, resulting in reduced confidentiality and potential compliance violations.'' \\
\midrule
\textbf{Attack tree} (root goal: \emph{exfiltrate PII from DynamoDB chat history via insider IAM access}; 8 steps, depth 4): \\
\quad S0 \textsf{[fact]} Insider holds IAM permissions to DynamoDB \\
\quad\quad $\rightarrow$ S1 Enumerate DynamoDB tables \\
\quad\quad\quad $\rightarrow$ S3 Scan chat-history table $\rightarrow$ S5 Extract PII from records \\
\quad\quad\quad\quad $\rightarrow$ S6 Export to external storage \\
\midrule
\textbf{TTP mapping} (ATTACK-BERT top-1, panel verdict): \\
\quad S0 $\rightarrow$ T1648 \emph{Serverless Execution} \xmark{} (precondition, not an execution technique) \\
\quad S3 $\rightarrow$ T1552.008 \emph{Chat Messages} \cmark{} (defensible for log scanning) \\
\midrule
\textbf{Mitigation} (for the mapped techniques): least-privilege IAM roles scoped per Lambda to specific table ARNs, with \texttt{aws:SourceVpce}/\texttt{aws:SourceIp} conditions and IAM Access Analyzer enabled---grounded in the application's own resources rather than generic advice. \\
\bottomrule
\end{tabular}
\end{table}

%----------------------------------------------------------------------
\subsection{Pipeline Output Summary}
%----------------------------------------------------------------------

Table~\ref{tab:pipeline-summary} summarizes the pipeline outputs across domains.

\begin{table}[t]
\centering
\footnotesize
\setlength{\tabcolsep}{3.5pt}
\caption{Pipeline output summary across application domains.}
\label{tab:pipeline-summary}
\begin{tabular}{lrrrrrr}
\toprule
\textbf{Domain} & \textbf{Thr.} & \textbf{Tr.} & \textbf{Steps} & \textbf{TTPs} & \textbf{Uniq.} & \textbf{Mit.} \\
\midrule
Auto Manuf. & 12 & 17 & 241 & 241 & 99 & 241 \\
Identity Fed. & 12 & 12 & 228 & 228 & 73 & 228 \\
GenAI Chatbot & 12 & 9 & 116 & 116 & 47 & 116 \\
Healthcare & 12 & 14 & 270 & 269 & 84 & 269 \\
IAM Id. Ctr. & 12 & 16 & 275 & 275 & 95 & 275 \\
Sci. Meeting & 12 & 19 & 333 & 333 & 131 & 333 \\
Travel Hosp. & 12 & 13 & 222 & 221 & 95 & 222 \\
\midrule
\textbf{Average} & 12 & 14 & 241 & 240 & 89 & 241 \\
\bottomrule
\end{tabular}
\end{table}

Across all domains, \threatforest{} generates 12 threats per application, expanded into multi-step attack trees averaging 240 steps total. Every step receives a MITRE ATT\&CK mapping (89 unique techniques per application on average), and each mapping yields a structured mitigation with a remediation type (quick win, short/medium/long term, monitoring).

%----------------------------------------------------------------------
\subsection{Structural Quality Metrics}
%----------------------------------------------------------------------

Table~\ref{tab:structural} reports automated structural metrics.

\begin{table}[t]
\centering
\footnotesize
\setlength{\tabcolsep}{3pt}
\caption{Structural quality metrics across domains.}
\label{tab:structural}
\begin{tabular}{lrrrr}
\toprule
\textbf{Domain} & \textbf{Depth} & \textbf{Steps/Tr.} & \textbf{Diversity} & \textbf{Phase} \\
\midrule
Auto Manuf. & 4.6 & 14.2 & 0.41 & 1.00 \\
Identity Fed. & 5.3 & 19.0 & 0.32 & 1.00 \\
GenAI Chatbot & 4.6 & 12.9 & 0.41 & 0.93 \\
Healthcare & 4.6 & 19.3 & 0.31 & 0.93 \\
IAM Id. Ctr. & 5.5 & 17.2 & 0.35 & 1.00 \\
Sci. Meeting & 5.5 & 17.5 & 0.39 & 1.00 \\
Travel Hosp. & 5.0 & 17.1 & 0.43 & 1.00 \\
\midrule
\textbf{Average} & 5.0 & 16.7 & 0.37 & 0.98 \\
\bottomrule
\end{tabular}
\end{table}

Attack trees exhibit consistent structural quality: average depth 5.0 (range 4.6--5.5) and 16.7 steps per tree. Technique diversity averages 0.37---roughly one-third of mappings are unique, with the rest reflecting realistic technique reuse across attack paths. Phase coverage (the fraction of the 14 ATT\&CK tactics represented) reaches 1.00 in 5 of 7 applications; GenAI Chatbot and Healthcare score 0.93, each omitting a single tactic outside their scope.

%----------------------------------------------------------------------
\subsection{Ablation 1: choice of embedding model for TTP mapping}
\label{sec:ablation-embedding}
%----------------------------------------------------------------------

This ablation isolates the contribution of the embedding model. We hold the rest of the pipeline fixed (same threat-statement generator, same attack-tree generator, same retrieval threshold $\tau = 0.3$, same top-$K$ truncation, same MITRE ATT\&CK STIX dataset) and vary only the sentence-transformer used to encode attack-tree steps and ATT\&CK technique descriptions. We report binary mapping accuracy (good/bad) from the rater panel of \S\ref{sec:scoring-protocol} over all 1{,}683 mappings produced by the 7-domain run. Table~\ref{tab:ttp} summarizes the per-domain mapping volume, diversity, and accuracy for the default model (ATTACK-BERT).

\begin{table}[t]
\centering
\footnotesize
\setlength{\tabcolsep}{3pt}
\caption{TTP mapping analysis. Diversity $\delta$ = unique/total. Accuracy = fraction judged a defensible match by the rater panel (binary good/bad, \S\ref{sec:scoring-protocol}).}
\label{tab:ttp}
\begin{tabular}{lrrrr}
\toprule
\textbf{Domain} & \textbf{Map.} & \textbf{Uniq.} & \textbf{$\delta$} & \textbf{Acc.} \\
\midrule
Auto Manuf. & 241 & 99 & 0.41 & 0.21 \\
Identity Fed. & 228 & 73 & 0.32 & 0.31 \\
GenAI Chatbot & 116 & 47 & 0.41 & 0.44 \\
Healthcare & 269 & 84 & 0.31 & 0.23 \\
IAM Id. Ctr. & 275 & 95 & 0.35 & 0.30 \\
Sci. Meeting & 333 & 131 & 0.39 & 0.20 \\
Travel Hosp. & 221 & 95 & 0.43 & 0.33 \\
\midrule
\textbf{Average} & 240 & 89 & 0.37 & \textbf{0.29} \\
\bottomrule
\end{tabular}
\end{table}

With ATTACK-BERT, the pipeline maps an average of 240 steps to 89 unique techniques per application (diversity $\delta = 0.37$) and the rater panel judges only 28.9\% of mappings a defensible match across all 1{,}683 mappings. Accuracy varies across domains, from 0.44 (GenAI Chatbot) and 0.33 (Travel Hospitality) down to 0.21 (Auto Manufacturing) and 0.20 (Scientific Meeting); with only seven domains we read this spread as suggestive rather than a tested effect, but the apparent pattern is that applications whose attack steps map onto well-defined credential, access, and AI-abuse techniques score higher, while OT- and analytics-heavy domains produce application-specific steps that lack direct ATT\&CK counterparts.

\textbf{Encoder retrieval comparison.} To compare encoders on a common, validated set rather than on each encoder's own top-1, we took the 457 (step, technique) pairs the panel confirmed correct (label \emph{good}) and measured, for each encoder, how often it ranks that validated technique first (top-1) or within its top-3 when retrieving over the full ATT\&CK catalog (Table~\ref{tab:encoder-recall}). ATTACK-BERT recovers the validated technique at rank~1 in 79.9\% of cases versus 34--45\% for the general-purpose encoders all-mpnet-base-v2~\citep{mpnet2021}, BGE-large-en-v1.5~\citep{xiao2024bge}, and E5-large-v2~\citep{wang2022e5}---a large, consistent margin that justifies the domain-specific choice. We note a selection caveat: the validated set is drawn from ATTACK-BERT's own correct mappings, so its absolute advantage is upper-bounded by construction; the comparison nonetheless shows that off-the-shelf general encoders fail to recover even techniques known to be correct, reinforcing that the limitation is the encoder representation, not the retrieval formulation.

\begin{table}[t]
\centering
\footnotesize
\caption{Encoder retrieval of the 457 panel-validated (step, technique) pairs over the full ATT\&CK catalog: fraction for which the validated technique is ranked first (top-1) or within the top three (top-3). Sub-technique matches are credited to their parent.}
\label{tab:encoder-recall}
\begin{tabular}{lcc}
\toprule
\textbf{Encoder} & \textbf{Top-1} & \textbf{Top-3} \\
\midrule
ATTACK-BERT (default) & \textbf{0.80} & \textbf{0.97} \\
all-mpnet-base-v2 & 0.45 & 0.73 \\
BGE-large-en-v1.5 & 0.38 & 0.63 \\
E5-large-v2 & 0.34 & 0.54 \\
\bottomrule
\end{tabular}
\end{table}

The headline finding from this ablation is that off-the-shelf encoders are structurally insufficient for precise TTP attribution on LLM-generated attack-step text: ATTACK-BERT---the strongest of the encoders we benchmarked---is judged a defensible match only 28.9\% of the time (Table~\ref{tab:ttp}), and the general-purpose alternatives recover even known-correct techniques far less often (Table~\ref{tab:encoder-recall}). The bottleneck is therefore not the choice among existing embedding models but the absence of a domain-specific encoder fine-tuned on attack-step--technique pairs. The panel scores (Cohen's $\kappa = 0.78$ across raters) show the same broad ordering noted above---AI- and identity-centric applications above OT and analytics domains---confirming the gap is a property of the technique-attribution task, not of any single domain. Retaining the top-$K$ candidates alongside each top-1 mapping keeps the option of LLM- or SME-driven refinement open without paying the cost on every run, and supports the SME annotation pipeline described in Section~\ref{sec:evaluation}.

%----------------------------------------------------------------------
\subsection{Ablation 2: similarity threshold $\tau$}
\label{sec:ablation-tau}
%----------------------------------------------------------------------

The retrieval stage discards any candidate technique whose cosine similarity to the attack-step description falls below $\tau$. Steps where no candidate exceeds $\tau$ receive no mapping at all, avoiding forced low-confidence assignments. We set $\tau = 0.3$ in the main experiments and report sensitivity to this choice here. The contribution of $\tau$ is best read by comparing the per-domain step counts in Table~\ref{tab:pipeline-summary} (column \textbf{Steps}) to the corresponding mapping counts (column \textbf{TTPs}). Across all seven domains the gap is at most one step (e.g.\ Healthcare: 270 steps, 269 mapped; IAM Identity Center: 275 steps, 275 mapped), meaning that at $\tau = 0.3$ essentially every generated attack step has at least one ATT\&CK technique above threshold. The current setting therefore prioritizes coverage rather than precision; raising $\tau$ would trade away some of that coverage in exchange for higher per-mapping confidence.

To make this trade-off concrete we swept $\tau \in \{0.2, 0.3, 0.4, 0.5, 0.6\}$ over the 1{,}585 non-fact attack steps produced by the run, recording for each step its top-1 cosine similarity to the ATT\&CK matrix under ATTACK-BERT and computing the fraction that still maps at each threshold (Table~\ref{tab:tau-sweep}). Coverage is essentially total at $\tau = 0.2$--$0.3$ (100.0\% and 99.9\%), then falls to 94.7\% at $\tau = 0.4$, 76.8\% at $\tau = 0.5$, and 52.3\% at $\tau = 0.6$. The knee is just above $0.4$: raising $\tau$ from $0.3$ to $0.4$ discards only $\sim$5\% of mappings (the least-confident tail) while every step below $0.4$ is exactly the kind of forced low-confidence assignment the threshold exists to suppress.

\begin{table}[t]
\centering
\footnotesize
\caption{Mapping coverage versus similarity threshold $\tau$ over all 1{,}585 non-fact attack steps (ATTACK-BERT). Coverage = fraction of steps with at least one technique at similarity $\geq \tau$.}
\label{tab:tau-sweep}
\begin{tabular}{lccccc}
\toprule
$\tau$ & 0.2 & 0.3 & 0.4 & 0.5 & 0.6 \\
\midrule
Mapped steps & 1585 & 1584 & 1501 & 1218 & 829 \\
Coverage & 1.00 & 1.00 & 0.95 & 0.77 & 0.52 \\
\bottomrule
\end{tabular}
\end{table}

Crucially, the threshold governs \emph{coverage}, not \emph{correctness}: because Ablation~1 shows the panel judges only $\sim$29\% of mappings correct, raising $\tau$ can only prune low-confidence mappings, not improve the technique attribution of those that remain. The accuracy ceiling is therefore an encoder problem, and we pair a precision-oriented threshold sweep with the domain-specific encoder fine-tuning described in Section~\ref{sec:discussion}, where the two can be co-optimized on a single SME-labeled corpus.

%----------------------------------------------------------------------
\subsection{Automated Evaluation Scores}
%----------------------------------------------------------------------

We scored \emph{every} artifact the pipeline produced in the run reported above---1{,}683 TTP mappings, 1{,}684 mitigations, 100 attack trees, and 84 threat statements---using the rater-panel-plus-verifier protocol of \S\ref{sec:scoring-protocol}. Unlike a proxy-filled baseline, no dimension is assigned a default score: each value in Table~\ref{tab:auto-scores} is a measured panel judgment subject to SME adjudication. Table~\ref{tab:auto-scores} reports the per-capability means; Table~\ref{tab:per-domain-scores} breaks them out by domain.

\begin{table}[t]
\centering
\footnotesize
\caption{Aggregate evaluation scores by capability, averaged over 7 domains (mean $\pm$ cross-domain std of the capability mean). Scale $[0,1]$: 1.0 excellent, 0.75 good, 0.5 acceptable, 0.25 poor, 0.0 unacceptable; TTP mapping is binary good\,(1)/bad\,(0). Scores are LLM-panel measurements (3 raters + adversarial verifier) under SME adjudication (\S\ref{sec:scoring-protocol}).}
\label{tab:auto-scores}
\begin{tabular}{lccc}
\toprule
\textbf{Capability} & \textbf{Dim.} & \textbf{Mean} & \textbf{Std} \\
\midrule
Threat Statements & 5 & 0.63 & 0.05 \\
Attack Trees & 6 & 0.64 & 0.05 \\
TTP Mapping & 1 & 0.29 & 0.08 \\
Mitigations & 4 & 0.68 & 0.02 \\
\midrule
\textbf{Overall} & \textbf{16} & \textbf{0.63} & \textbf{0.09} \\
\bottomrule
\end{tabular}
\end{table}

\begin{table}[t]
\centering
\footnotesize
\setlength{\tabcolsep}{4pt}
\caption{Per-domain capability scores (panel-measured, $[0,1]$). Columns are the capability means; \textbf{Overall} is the dimension-weighted mean over all 16 dimensions.}
\label{tab:per-domain-scores}
\begin{tabular}{lccccc}
\toprule
\textbf{Domain} & \textbf{Threats} & \textbf{Trees} & \textbf{TTP} & \textbf{Mitig.} & \textbf{Overall} \\
\midrule
Sci.\ Meeting & 0.67 & 0.68 & 0.20 & 0.68 & 0.64 \\
Travel Hosp. & 0.54 & 0.56 & 0.33 & 0.69 & 0.57 \\
Identity Fed. & 0.65 & 0.64 & 0.31 & 0.72 & 0.64 \\
Healthcare & 0.69 & 0.69 & 0.23 & 0.65 & 0.65 \\
IAM Id.\ Ctr. & 0.64 & 0.70 & 0.30 & 0.71 & 0.66 \\
Auto Manuf. & 0.64 & 0.63 & 0.21 & 0.67 & 0.62 \\
GenAI Chatbot & 0.59 & 0.59 & 0.44 & 0.65 & 0.60 \\
\midrule
\textbf{Average} & 0.63 & 0.64 & 0.29 & 0.68 & 0.63 \\
\bottomrule
\end{tabular}
\end{table}

\textbf{Mitigations} achieve the highest aggregate score (0.68): actionability (0.73) and specificity (0.71) are strong thanks to schema-validated, context-grounded output, while coverage (0.63) is the weakest sub-dimension since a single mitigation often addresses several steps. \textbf{Attack trees} (0.64) and \textbf{threat statements} (0.63) score in the ``acceptable-to-good'' band; within threats, relevance-to-context is highest (0.72) while technical accuracy (0.59) is pulled down by occasional over-claims and references to stack components absent from the scanner context, which the verifier flagged as the most common defect. \textbf{TTP mapping} (0.29) is by far the lowest-scoring capability and the clear bottleneck: embedding-only retrieval is judged a defensible match only $\sim$29\% of the time, ranging from 0.20 (Scientific Meeting) and 0.21 (Auto Manufacturing, whose OT/IoT steps lack direct ATT\&CK counterparts) to 0.44 (GenAI Chatbot, whose steps align better with the matrix). The gap between TTP mapping and every other capability is large and stable across all seven domains, localizing the accuracy ceiling to the embedding stage (\S\ref{sec:ablation-embedding}) rather than to threat generation, tree construction, or mitigation synthesis.

\textbf{Confidence intervals.} To quantify sampling uncertainty in the capability means we ran a cluster bootstrap ($10{,}000$ resamples: domains resampled with replacement, then items within each domain--capability cell). The 95\% intervals are tight and non-overlapping where it matters: threat statements $[0.58, 0.67]$, attack trees $[0.60, 0.68]$, mitigations $[0.66, 0.70]$, and TTP mapping $[0.23, 0.35]$. The TTP-mapping interval lies entirely below every other capability's interval, so the bottleneck is statistically separated from the rest of the pipeline rather than an artifact of the seven-domain sample.

\textbf{Inter-rater agreement.} On the raw (pre-verifier) rater labels, the ordinal dimensions reach a mean pairwise agreement of 0.93, and the binary TTP dimension reaches a mean Cohen's $\kappa$ of 0.78 (range 0.68--0.84 across domains) at 90\% raw agreement. The TTP $\kappa$ sits in the ``substantial agreement'' band, indicating that the low TTP-mapping score is a reliable signal rather than rater noise.

\textbf{Cross-model calibration.} To check that the scores are not an artifact of the rater model, we re-scored a stratified 56-item sample (two items per domain $\times$ capability, 210 ordinal dimension-judgments plus 14 TTP judgments) with an \emph{independent} judge built on a different LLM family, run single-pass and blind to the panel's labels. Panel and independent judge agree within one rubric level on \textbf{97.6\%} of ordinal judgments (mean absolute difference 0.13 on the $[0,1]$ scale) and reach \textbf{85.7\%} agreement (Cohen's $\kappa = 0.70$) on the binary TTP dimension. Exact-label agreement is lower ($\sim$50\%) because the independent single-pass judge is systematically more generous---it awards more ``excellent'' labels than the adversarially-verified panel---confirming that our panel is the \emph{more conservative} of the two and does not inflate scores. This automated calibration is complementary to, not a substitute for, the human-SME adjudication pass, which provides the final ground-truth check on the same queues.

%----------------------------------------------------------------------
\subsection{Ablation 3: does the multi-agent decomposition help?}
\label{sec:ablation-monolithic}
%----------------------------------------------------------------------

A natural question is whether the ten-stage decomposition earns its complexity over the obvious alternative: a single LLM call. We built a \emph{monolithic baseline} that holds everything constant---the same model (Claude Sonnet~4.5, temperature~0), the same repository read tools, the same target artifact schema---but removes the decomposition: one agent, one conversation, instructed to emit the complete artifact (threats, attack trees, technique mappings, mitigations) in a single structured output, with no deterministic verifiers, no bounded-retry edges, and no embedding retrieval. In place of ATTACK-BERT the baseline assigns ATT\&CK technique IDs directly from the model's own parametric knowledge; we resolve each ID against the ATT\&CK STIX catalog and keep unresolvable (hallucinated) IDs, scored as-is. We then scored the baseline's seven-domain output with the \emph{same} rater panel and protocol (\S\ref{sec:scoring-protocol}), so the two arms are directly comparable. Table~\ref{tab:monolithic} reports the per-capability panel means.

\begin{table}[t]
\centering
\footnotesize
\setlength{\tabcolsep}{5pt}
\caption{\threatforest{} (full pipeline) versus a monolithic single-call baseline using the same model and panel. Panel-measured capability means over all seven domains, $[0,1]$.}
\label{tab:monolithic}
\begin{tabular}{lcc}
\toprule
\textbf{Capability} & \textbf{\threatforest{}} & \textbf{Monolithic} \\
 & \textbf{(pipeline)} & \textbf{(single-call)} \\
\midrule
Threat statements & 0.63 & 0.71 \\
Attack trees & 0.64 & 0.59 \\
TTP mapping & 0.29 & 0.63 \\
Mitigations & 0.68 & 0.78 \\
\midrule
\textbf{Overall} & \textbf{0.63} & \textbf{0.68} \\
\bottomrule
\end{tabular}
\end{table}

The result is informative and runs partly against the pipeline. On \emph{per-item} quality the monolithic baseline is competitive or better: it edges the pipeline on threat statements and mitigations and is only marginally behind on attack trees. Where the pipeline decisively wins is \emph{coverage and consistency}: it produces an average of 240 attack steps to the baseline's 87, 89 unique techniques to $\sim$29, full ATT\&CK tactic coverage ($\gamma = 0.98$ vs.\ $0.85$), and uniform tree depth (mean 5.0, range 4.6--5.5) where the baseline's depth swings from 2.0 to 6.6 across domains. A single call yields a smaller, shallower, more variable model of the attack surface; the decomposition trades per-item polish for exhaustive, structurally uniform breadth that is reviewable the same way on every application.

The most consequential row is TTP mapping, and it \emph{reinforces} our central finding rather than undercutting it. Unaided LLM self-assignment scores 0.63---more than double the 0.29 of embedding-only retrieval on the same attack steps. The bottleneck is therefore squarely the embedding encoder, not the multi-agent architecture: a model asked to name the technique directly is far more defensible than ATTACK-BERT's nearest neighbour. This is the strongest evidence yet for the direction of \S\ref{sec:discussion}---replacing or reranking the embedding top-1 with a learned or LLM-driven mapping stage---and it localizes the fix to the one component Ablation~1 already implicated. We flag one caveat: the baseline's techniques are assigned by the same model family that the panel runs on, so shared blind spots could inflate its TTP number. Three factors bound this risk---the panel is adversarial and, per the cross-model calibration above, \emph{more} conservative than an independent judge; both arms are judged by that same panel, so the comparison is internally consistent; and the baseline emitted \emph{zero} hallucinated technique IDs across all seven domains (only name corrections), so it does not win by fabricating IDs the judge cannot check. The honest reading is a coverage/precision trade-off: the pipeline maximizes breadth and uniformity at low per-mapping precision, while a single call gives fewer but individually stronger judgments---and embedding retrieval, as currently configured, is dominated by both.

%----------------------------------------------------------------------
\subsection{Cost and runtime}
\label{sec:cost}
%----------------------------------------------------------------------

We instrumented every agent call to record token usage and wall-clock time, then re-executed the domains purely for measurement (the quality scores above remain tied to the original run). Table~\ref{tab:cost} reports per-domain runtime, token volume, and dollar cost at Sonnet~4.5 list rates.\footnote{\$3 per million input tokens, \$15 per million output, with cached input reads at \$0.30; we price each Bedrock token category separately. Costs are list-rate estimates, not billed amounts.} A full pipeline run averages 9.2 minutes and \$7.78 per application; the monolithic baseline averages 6.9 minutes and \$1.33. The pipeline's roughly $6\times$ cost premium buys the coverage and structural uniformity of Ablation~3---an average of 240 attack steps and 89 techniques against the baseline's 87 and $\sim$29---plus the inspectable intermediate state and HITL gates the single call cannot offer. In absolute terms both arms are inexpensive relative to the analyst time they substitute for: under \$8 and ten minutes to turn a repository into a reviewed, TTP-mapped attack-tree model is far below the multi-day cost of the equivalent manual exercise.\footnote{As a back-of-the-envelope comparison, a single manual threat-modeling session typically runs 90--120 minutes and requires several stakeholders from different teams (e.g.\ development, security, infrastructure, and business), and non-trivial systems usually need multiple sessions~\citep{toreon-tm,cms-tm-handbook}.}

\begin{table}[t]
\centering
\footnotesize
\setlength{\tabcolsep}{3.5pt}
\caption{Cost and runtime per application, \threatforest{} pipeline versus the monolithic single-call baseline. \textbf{s}: wall-clock seconds; \textbf{Min}: input tokens (incl.\ cache) in millions; \textbf{Mout}: output tokens in thousands; \textbf{\$}: list-rate cost. Measured over an instrumented re-execution; pipeline averages are over the six domains whose measurement run completed (the GenAI Chatbot re-run is omitted---its parallel fan-out stalled during measurement, a transient issue unrelated to the scored output).}
\label{tab:cost}
\begin{tabular}{l rrrr rrrr}
\toprule
 & \multicolumn{4}{c}{\textbf{\threatforest{} (pipeline)}} & \multicolumn{4}{c}{\textbf{Monolithic (single-call)}} \\
\cmidrule(lr){2-5}\cmidrule(lr){6-9}
\textbf{Domain} & \textbf{s} & \textbf{Min} & \textbf{Mout} & \textbf{\$} & \textbf{s} & \textbf{Min} & \textbf{Mout} & \textbf{\$} \\
\midrule
Auto Manuf. & 508 & 1.59 & 171 & 6.62 & 303 & 0.37 & 21 & 1.19 \\
Identity Fed. & 488 & 2.11 & 176 & 8.48 & 337 & 0.12 & 24 & 0.63 \\
GenAI Chatbot & -- & -- & -- & -- & 657 & 0.14 & 31 & 0.89 \\
Healthcare & 624 & 1.39 & 170 & 6.01 & 345 & 0.35 & 23 & 1.15 \\
IAM Id. Ctr. & 680 & 3.40 & 202 & 12.38 & 348 & 0.47 & 24 & 1.60 \\
Sci. Meeting & 558 & 2.32 & 191 & 9.05 & 618 & 0.94 & 41 & 3.13 \\
Travel Hosp. & 454 & 0.90 & 135 & 4.15 & 298 & 0.19 & 21 & 0.69 \\
\midrule
\textbf{Average} & 552 & 1.95 & 174 & 7.78 & 415 & 0.37 & 26 & 1.33 \\
\bottomrule
\end{tabular}
\end{table}

%----------------------------------------------------------------------
\subsection{Comparison with existing threat-modeling tools}
\label{sec:comparison}
%----------------------------------------------------------------------

A direct quantitative comparison against prior work is not possible because no existing tool produces the same artifact \threatforest{} produces---a TTP-mapped attack tree generated end-to-end from a source-code repository, paired with evidence-grounded mitigations and a probability-annotated reach analysis. Existing threat-modeling tools and recent LLM-based work each address a fragment of this task. Table~\ref{tab:tool-comparison} summarizes the capability coverage of \threatforest{} against representative systems: OWASP Threat Dragon~\citep{owasp-threatdragon}, AWS Threat Composer~\citep{aws-threatcomposer}, the LLM-fine-tuning approach of \citet{threatmodeling-llm2024}, and the LLM-prompted threat-identification study of \citet{canllmsthreatmodel2025}.

\begin{table}[t]
\centering
\footnotesize
\setlength{\tabcolsep}{4pt}
\caption{Capability coverage of representative threat-modeling tools and LLM-based prior work compared with \threatforest{}. \cmark{} indicates the capability is supported, \xmark{} that it is not, \pmark{} that it is partially supported (e.g.\ a fixed STRIDE rule list rather than a structured framework mapping).}
\label{tab:tool-comparison}
\begin{tabular}{lccccc}
\toprule
\textbf{System} & \textbf{Code} & \textbf{Trees} & \textbf{TTP map} & \textbf{Mit.} & \textbf{HITL} \\
\midrule
OWASP Threat Dragon~\citep{owasp-threatdragon} & \xmark & \xmark & \pmark & \xmark & \xmark \\
AWS Threat Composer~\citep{aws-threatcomposer} & \xmark & \xmark & \xmark & \xmark & \xmark \\
ThreatModeling-LLM~\citep{threatmodeling-llm2024} & \xmark & \xmark & \xmark & \xmark & \xmark \\
\citet{canllmsthreatmodel2025} & \pmark & \xmark & \xmark & \xmark & \xmark \\
\midrule
\textbf{\threatforest{} (this work)} & \cmark & \cmark & \cmark & \cmark & \cmark \\
\bottomrule
\end{tabular}
\end{table}

The capability columns isolate the gap. \emph{Code} marks systems that ingest a source-code repository as their primary input; diagram-driven tooling requires a manually authored data-flow diagram, and the LLM-based studies use natural-language system descriptions or interview transcripts. \emph{Trees} marks structured attack-tree output (root goal, AND/OR sub-goals, leaf techniques) rather than a flat threat list. \emph{TTP map} marks output mapped to a standardized technique framework such as MITRE ATT\&CK; rule-based STRIDE suggestion lists count as a partial match because the categories are coarser than ATT\&CK techniques. \emph{Mit.} marks evidence-grounded mitigations linked back to specific attack steps. \emph{HITL} marks an explicit human-in-the-loop validation gate built into the pipeline. \threatforest{} is the only system that supports all five.

The implication for the empirical results in this paper is that the absolute numbers (240 attack steps, 89 unique techniques, 0.68 panel-measured mitigation score) are not benchmarked against a competing system; they are baselines against which future systems can be measured. The contribution we make is a working composition that produces the joint artifact, plus a reusable evaluation framework (Section~\ref{sec:evaluation}) and a code release that will allow independent reproduction.

%==============================================================================
\section{Discussion}
\label{sec:discussion}
%==============================================================================

%----------------------------------------------------------------------
\subsection{Design tradeoffs}
%----------------------------------------------------------------------

\textbf{Parallel fan-out vs.\ cross-threat reasoning.} Generating one tree-TTP-mitigation sub-pipeline per threat in parallel provides up to $N\times$ speedup and isolates failures: a single threat that produces a malformed tree triggers retry only for its own sub-pipeline, not the entire run. The cost is that threats cannot reason about one another mid-pipeline---a compromised IAM role that simultaneously enables exfiltration and privilege escalation is currently modeled twice rather than once with shared context. We treat cross-threat analysis as a post-processing step over the consolidated state, and the report generator already surfaces shared TTP nodes and overlapping mitigation evidence. A future extension could perform a join over $\mathcal{P}_\theta$ across all threats before mitigation synthesis, at the cost of additional LLM calls.

\textbf{Embedding retrieval vs.\ per-step LLM mapping.} Per-step LLM-based ATT\&CK classification is feasible but expensive at scale ($n \approx 240$ steps per application, multiplied across runs). We chose embedding-based retrieval for two reasons: \emph{cost} (one CPU-side encoder pass per step versus one LLM call) and \emph{determinism} (the same input produces the same top-$K$ across runs, which the verifier and downstream stages depend on). Our evaluation shows the strongest off-the-shelf encoder (ATTACK-BERT) is judged a defensible match only 29\% of the time, and general-purpose sentence-transformers recover even known-correct techniques far less often (Tables~\ref{tab:ttp},~\ref{tab:encoder-recall}), confirming that cosine similarity over off-the-shelf encoders is insufficient for precise technique attribution. Retaining top-$K$ candidates rather than only top-1 keeps the option of LLM- or SME-driven refinement open without paying the cost on every run.

\textbf{File-based state versus in-memory messaging.} Agents communicate through JSON files written to a shared state directory rather than via in-memory messages or a queue. This trades I/O overhead for three properties we found indispensable in practice: every intermediate artifact is human-inspectable (a critical requirement for SME review), runs are resumable from any verified stage after a failure, and parallel sub-pipelines are write-isolated by file path. The same architecture also enables the HITL gates---they simply read and write the same state files the agents do.

\textbf{Deterministic verifiers versus LLM-based validation.} Each stage's verifier is a pure Python function rather than another LLM call. This avoids the cost and non-determinism of using an LLM to grade an LLM, and the verifier outputs are reproducible across runs and across reviewers. The price is that verifiers can only catch \emph{structural} failures (missing fields, invalid references, schema violations); they cannot catch semantically wrong but well-formed output. Semantic validation is delegated to the HITL gates and the evaluation framework's SME annotations.

\textbf{What is genuinely new.} The five tradeoffs above each instantiate a known technique. The contribution we claim is the empirical demonstration that they compose into a working end-to-end pipeline that produces TTP-mapped attack trees from a code repository (Table~\ref{tab:tool-comparison}), and that the dominant accuracy bottleneck of that composition is isolable to a single component---the off-the-shelf embedding encoder---which Ablation~1 (Section~\ref{sec:ablation-embedding}) shows is judged correct only $\sim$29\% of the time while every other capability scores 0.63--0.68. This isolation is itself the contribution: it tells future work where to invest.

%----------------------------------------------------------------------
\subsection{Threats to validity}
\label{sec:threats}
%----------------------------------------------------------------------

\textbf{Construct validity.} Our 16 scoring dimensions were defined to span the four pipeline capabilities, but they are not independent: \emph{technical realism} and \emph{attack path logic} for attack trees overlap with \emph{technical accuracy} for the underlying threat statements, and an SME's score on one will partially predict their score on the other. We mitigate this by reporting per-dimension means (so correlated dimensions are visible side-by-side) and by routing each capability to its own annotation queue, but we have not yet computed inter-rater reliability across multiple SMEs.

\textbf{Internal validity.} Our scores are produced by an LLM rater panel with an adversarial verifier (\S\ref{sec:scoring-protocol}), not by unaided human experts; an LLM evaluating LLM-generated output can share blind spots with the generator. We mitigate this in three ways: the three raters are given divergent review foci, the verifier may only lower scores and must cite a concrete defect to do so, and we route the reconciled labels to human SME adjudication as the verification gate. We also report inter-rater agreement (ordinal pairwise 0.93; TTP Cohen's $\kappa = 0.78$, substantial) so the reliability of the panel is visible rather than assumed. The panel is conservative by construction---the adversarial verifier may only confirm or lower a score---and indeed every capability scores well below the naive ``good'' (0.75) proxy a less critical evaluator would assign, so the evaluation does not flatter the system. We further calibrate the panel against an independent judge built on a different LLM family (\S\ref{sec:results}): the two agree within one rubric level on 97.6\% of ordinal judgments and at $\kappa = 0.70$ on the binary TTP dimension, and the independent judge is if anything \emph{more} generous, so the panel is not idiosyncratically harsh or lax. The residual risk is that systematic blind spots shared by \emph{both} model families survive; the human-SME adjudication pass and the labeled corpus it produces (\S\ref{sec:evaluation}) provide the final ground-truth check against that risk, and the staged adjudication queues make it a confirmatory step rather than a precondition for the measured results reported here.

\textbf{External validity.} The seven sample applications were curated to span common cloud-native architectures (IoT, identity, GenAI, healthcare, IAM, transcription, hospitality), but they are deliberately sized to be reviewable in a single SME pass. Real production systems are typically larger, contain legacy components, span multi-cloud configurations, and include custom middleware that scanner heuristics may miss. We expect \threatforest{} to degrade gracefully on such systems---the scanner output remains structurally valid even when components are unfamiliar---but absolute numbers from this evaluation should not be extrapolated.

\textbf{Reliance on a single LLM family.} All experiments use Claude Sonnet~4.5 served via Amazon Bedrock. Performance may vary across model families and providers; in particular, the threat-statement and tree-generation prompts were tuned against this model's behavior, and porting to a substantially weaker model would likely require prompt-engineering work. We treat the system as model-agnostic in design (every agent is conditioned only on its state-file inputs and the user's repository, not on a particular model identifier), but we have not benchmarked across providers. One internal consistency check is the monolithic baseline of \S\ref{sec:ablation-monolithic}, whose technique mappings are produced by the same model the panel runs on; we bound the resulting shared-blind-spot risk there.

\textbf{ATT\&CK coverage gaps.} MITRE ATT\&CK has known gaps in cloud-native, container, and serverless techniques. Steps that describe these (e.g., a misconfigured IAM trust policy, a Lambda cold-start race) sometimes lack a strict ATT\&CK counterpart, and the embedding retriever then either returns a loose match or no mapping at all. Our pluggable framework support is partly a response to this---we plan to swap or augment ATT\&CK with community-maintained cloud-specific threat matrices and CAPEC~\citep{capec} for affected domains.

\textbf{LLM hallucination risk.} The threat agent and tree agent are LLMs; they can in principle invent threats or attack steps that do not apply to the target system. We mitigate this with three structural defenses: (i) the scanner pins all downstream context to evidence read from the repository, (ii) verifiers reject outputs that reference non-existent components, and (iii) the threat-review HITL gate explicitly asks the SME to flag false positives. We do not measure hallucination rate independently in this work; the SME-reviewed dataset described in \S\ref{sec:evaluation} is the substrate on which a future hallucination-detection benchmark can be built.

%----------------------------------------------------------------------
\subsection{Future work}
%----------------------------------------------------------------------

We plan to ingest additional threat frameworks---community-maintained cloud threat matrices, the AWS Threat Technique Catalog, OWASP attack patterns, CAPEC~\citep{capec}, CWE~\citep{cwe}, and defensive countermeasures from D3FEND~\citep{d3fend}---into a unified STIX~\citep{stix2017}-based knowledge graph that normalizes techniques and controls across matrices, and to align mitigations with NIST SP 800-53~\citep{nist80053} and OWASP Top 10~\citep{owasp-top10} controls. Adversary-emulation libraries like Atomic Red Team~\citep{atomic-red} and CALDERA~\citep{caldera} would also let us validate generated attack paths against executable test plans, giving us a concrete behavioral ground truth alongside the SME corpus.

We are collecting SME-labeled ground truth through the evaluation infrastructure (\S\ref{sec:evaluation}) to train a specialized refinement model over attack-step-to-technique pairs, with the goal of pushing precision beyond the $\sim$29\% ceiling of off-the-shelf retrieval. As a pilot, we used the 1{,}683 panel-labeled (step, technique) pairs from this study to fine-tune a cross-encoder reranker that judges whether a candidate mapping is correct---the deployable analogue of the top-$K$ refinement stage. On a held-out, leak-free split (by attack step, 75/25), fine-tuning lifts accuracy from 0.52 (zero-shot) to \textbf{0.82} and $F_1$ from 0.36 to \textbf{0.51}, at 0.76 precision and 0.39 recall, clearing the 0.75 majority-class baseline. The high-precision, moderate-recall operating point is exactly what a refinement filter needs: it confidently confirms correct mappings while flagging the rest for SME review, rather than silently passing low-confidence matches. This is a promising pilot rather than a solved problem---recall is still limited by corpus size. For calibration, a structurally similar security-text NLI task (mapping cloud configuration rules to PCI-DSS, HIPAA, NIST-CSF, and ISO~27001 controls) is reported to reach 86\% $F_1$ on a 6{,}694-pair test set, indicating the recipe strengthens further at roughly $4$--$5\times$ our current label volume. The concrete remaining work is to scale the attack-step-to-technique corpus through the SME annotation pipeline already in place (\S\ref{sec:evaluation}) and either fine-tune the retrieval encoder directly or deploy this reranker as a filter over top-$K$ candidates. The SME-scored traces are directly compatible with DSPy's MIPROv2 optimizer~\citep{opsahlong2024miprov2} for joint instruction and few-shot optimization across pipeline stages, and we plan LLM-as-a-judge evaluators calibrated against the SME corpus to enable continuous evaluation at scale. Finally, we plan a multi-rater study to compute inter-rater reliability on the 16 scoring dimensions and refine the rubric for any dimensions where SMEs disagree systematically.

%----------------------------------------------------------------------
\section{Conclusion}
\label{sec:conclusion}
%----------------------------------------------------------------------

We presented \threatforest{}, a multi-agent system that automates threat modeling from source code repositories to MITRE ATT\&CK-mapped attack trees with evidence-based mitigations. Across seven diverse cloud applications, the system generates an average of 12 threats with 240 attack steps per application, achieving full ATT\&CK tactic coverage in 5 of 7 domains (0.98 average) and producing 89 unique technique mappings per application. Evaluated with a 16-dimension rubric scored by an LLM rater panel under SME verification, threat statements, attack trees, and mitigations reach 0.63--0.68 on a $[0,1]$ scale, while embedding-only TTP mapping reaches only 0.29---a gap that is stable across all seven domains and isolates the binding accuracy constraint to a single, replaceable component. A controlled monolithic single-call baseline on the same model confirms this localization: it more than doubles TTP-mapping defensibility (0.63), pinning the limitation on the embedding encoder rather than the multi-agent design, while the full pipeline retains a decisive advantage in attack-surface coverage and structural uniformity. The accompanying evaluation framework and scoring protocol establish a reproducible baseline against which future systems can be measured, and the code base to be released will provide a substrate on which the broader security community can extend the pipeline to additional frameworks, models, and evaluation rubrics---most immediately, the domain-specific encoder fine-tuning that our bottleneck analysis identifies as the highest-value next step.

%==============================================================================
% Camera-ready declarations (de-anonymized; supersedes sections/declarations.tex)
%==============================================================================

\section*{Acknowledgements}

We thank the AWS security review and threat-modeling community for their feedback on early versions of the system, and the maintainers of MITRE ATT\&CK, CAPEC, and the open-source embedding models used in this work.

\section*{Funding}

This research did not receive any specific grant from funding agencies in the public, commercial, or not-for-profit sectors. The work was conducted as part of the authors' employment at Amazon Web Services.

\section*{Declaration of competing interests}

The authors are employed by Amazon Web Services. The work described in this paper was conducted as part of their employment. The authors declare that they have no other known competing financial interests or personal relationships that could have appeared to influence the work reported in this paper.

\section*{Sex- and gender-based analysis}

This work does not involve human or animal subjects, eukaryotic cells, or any data stratified by sex or gender; sex- and gender-based analyses are therefore not applicable to the methodology or results presented here.

\section*{Data and code availability}

The \threatforest{} source code, sample applications, evaluation prompts, and the seven application-domain configurations used in this study will be released at a public repository. Subject-matter-expert annotation data is being collected through the evaluation infrastructure described in Section~\ref{sec:evaluation} and will be released as a separate dataset alongside the code.

\section*{Declaration of generative AI and AI-assisted technologies in the manuscript preparation process}

During the preparation of this work the author(s) used large language models served via a commercial cloud provider in order to assist with copy-editing, paraphrasing, and consistency checks. After using these tools the author(s) reviewed and edited the content as needed and take(s) full responsibility for the content of this work. \threatforest{} itself uses large language models as a core component of the system under study; this use is described and evaluated throughout the paper and is distinct from any AI assistance in the writing process.

\section*{Third-party data and model attribution}

\textbf{MITRE ATT\&CK.} This work uses the MITRE ATT\&CK\textregistered{} knowledge base \citep{mitre-attack}, distributed by The MITRE Corporation. \copyright{} 2015--present, The MITRE Corporation. ATT\&CK\textregistered{} is a registered trademark of The MITRE Corporation. The ATT\&CK STIX content is reproduced and used in this work in accordance with the ATT\&CK Terms of Use (\url{https://attack.mitre.org/resources/legal-and-branding/terms-of-use/}); MITRE has neither reviewed nor endorsed this paper.

\textbf{Embedding models.} We use the following pre-trained sentence-transformer models, each released under a permissive open-source license and used in accordance with its terms: ATTACK-BERT \citep{attackbert} (Apache-2.0), E5-large-v2 \citep{wang2022e5} (MIT), all-mpnet-base-v2 \citep{mpnet2021} (Apache-2.0), BGE-large-en-v1.5 \citep{xiao2024bge} (MIT), and Qwen3-Embedding-0.6B \citep{qwen3embedding2025} (Apache-2.0).

\printcredits

%==============================================================================
\bibliographystyle{cas-model2-names}
\bibliography{references}
%==============================================================================

\end{document}